\documentstyle[pre,aps,floats,epsf,amssymb]{revtex}
\input epsf
\begin{document}
\draft

%\twocolumn[\hsize\textwidth\columnwidth\hsize\csname@twocolumnfalse%
%\endcsname
\title{
Steady-states and kinetics of ordering 
in bus-route models:  
connection with the Nagel-Schreckenberg model}
%\vspace{1cm}

\author{Debashish Chowdhury${\dag}$ and Rashmi C. Desai$^*$}

\address{ Department of Physics, University of Toronto,
Toronto, ON, M5S 1A7  Canada }

\date{\today}
\maketitle

%\documentstyle{article}
%\documentstyle[12pt]{article}
%\begin{document}
%\topmargin -1cm 

%\vspace{1cm}

%\centerline{\bf Steady-states and kinetics of ordering}
%\centerline{\bf in a bus-route model:}
%\centerline{\bf connection with the Nagel-Schreckenberg model}
%\vspace{1cm}
%\centerline{\bf Debashish Chowdhury$^1$ and Rashmi C. Desai}
%%\centerline{Department of Physics, University of Toronto} 
%\centerline{Toronto, ON, M5S 1A7  Canada}
%\vspace{1cm}
%\end{center} 
%\renewcommand{\baselinestretch}{2}\huge\normalsize
\begin{abstract}

A Bus Route Model (BRM) can be defined on a 
one-dimensional lattice, where buses are represented by 
"particles" that are driven forward from one site to the 
next with each site representing a bus stop. We replace 
the random sequential updating rules in an earlier BRM by parallel 
updating rules. In order to elucidate the connection between 
the BRM with parallel updating (BRMPU) and the Nagel-Schreckenberg 
(NaSch) model, we propose two alternative extensions of the NaSch 
model with space-/time-dependent hopping rates. Approximating 
the BRMPU as a generalization of the NaSch model, we calculate 
analytically the steady-state distribution of the {\it time 
headways} (TH) which are defined as the time intervals between 
the departures (or arrivals) of two successive particles (i.e., 
buses) recorded by a detector placed at a fixed site (i.e., bus 
stop) on the model route. We compare these TH distributions 
with the corresponding results of our computer simulations of 
the BRMPU, as well as with the data from the simulation of the 
two extended NaSch models. We also investigate interesting kinetic 
properties exhibited by the BRMPU during its time evolution 
from random initial states towards its steady-states.

\end{abstract}

%\vspace{0.5cm}
\noindent
---------------------------------------------------------------------------------


\noindent $^\dag$~Permanent address: Physics Department, I.I.T.,
Kanpur  208016, INDIA

\noindent $^*$~email: desai@physics.utoronto.ca

\newpage
%\vfill\eject 
%\renewcommand{\baselinestretch}{2}\huge\normalsize

\section{Introduction}

Systems of interacting particles driven far from equilibrium are 
of current interest in statistical physics \cite{sz,gs,spohn,vp,cursci}. 
Microscopic models of such systems often capture some aspects of 
vehicular traffic. In such "particle-hopping" models of vehicular 
traffic the particles represent vehicles and the nature of the 
interactions among these particles is determined by the manner in 
which the vehicles influence the motion of each other 
\cite{chowpr,proc1,proc2}. 
The dynamics of these models are often formulated in terms of ~
"update rules" using the language of cellular automata (CA) 
\cite{wolfram}. For example, the Nagel-Schreckenberg (NaSch) model 
\cite{ns,ssni} is the most popular minimal CA model of vehicular 
traffic on highways while, to our knowledge, the first CA model 
of city traffic was developed by Biham, Middleton and Levin 
\cite{bml}. The results obtained for these models, using the 
techniques of statistical mechanics, are not only of fundamental 
intetest for understanding truly nonequilibrium phenomena but 
may also find practical use in traffic science and engineering. 
\cite{may,leutz,nagel}. Among such results are the time-headway and
distance-headway distributions.
The {\it time-headway}({\bf TH}) is defined as the 
time interval between the departures (or arrivals) of two successive 
vehicles recorded by a detector placed at a fixed position on the 
route while the distance between the successive vehicles can be 
defined as the corresponding distance-headway ({\bf DH}). The 
distributions of TH and DH not only contain detailed informations 
on the nature of the spatio-temporal organization of the vehicles 
but are also of practical interest to traffic engineers because 
larger headways provide greater margins of safety whereas higher 
capacities of the highway require smaller headways. 

In a $1998$ paper, O'Loan et al \cite{evans} have developed a one-
dimensional lattice model of bus-route where the buses are 
represented by particles which move from one site to the next; 
each site of this model represents a bus stop along the route. 
The motion of the buses in this bus route model with random 
sequential updating (BRMRSU) is strongly influenced by the 
passengers waiting at the bus stops. The BRMRSU model may be 
viewed as a generalization of a simple particle-hopping model, 
namely, the totally asymmetric simple exclusion process (TASEP)  
by coupling the dynamics of the particles to another new 
variable which represents the presence (or absence) of passengers 
waiting at the bus stops. The bus route model in \cite{evans} does
not deal with overcrowded buses; it implicitly assumes 
that either the buses have infinite capacity or that the passenger 
arrival rate is slow enough to avoid overcrowding.

The BRMRSU exhibits a Bose-Einstein-condensation-like phenomenon 
which has been observed earlier in the TASEP and in the NaSch model 
when quenched random hopping rates are associated with 
the particles \cite{evansbe,krug,ktitarev}. However, unlike the 
stable Bose-Einstein-condensed states observed at sufficiently low 
densities in the TASEP (and in the NaSch model) with random hopping 
rates, those in the BRMRSU are metastable. The main 
characteristic of the spatially-inhomogeneous Bose-Einstein-condensed 
state is the existence of a macroscopically long gap in front of 
a cluster of vehicles led by the slowest one. In finite systems, 
for small $\lambda$ ($\lambda$ is the rate of passenger arrival at 
a bus stop), the bus clusters (or, equivalently, the gaps 
between clusters) in the BRMRSU exhibit interesting coarsening 
phenomena as the system evolves from a random initial state.
O'Loan et al. \cite{evans} find that, after sufficiently long time, 
the typical size of the large gaps 
in the system grows with time $t$ according to a power growth law $t^{1/2}$. 

In this paper we use a BRM with {\it parallel updating} ({\bf 
BRMPU}) which is obtained from the BRMRSU \cite{evans} by 
replacing the random sequential updating rule with parallel 
updating, with the aim of relating it with the NaSch model where
updating is done in parallel. We also propose here two extensions of 
the NaSch model (from now onwards referred to as the models {\bf Y} 
and {\bf Z}) by replacing the constant hopping rates with two  
different time-/space-dependent hopping rates which we shall 
specify explicitly in section II. We have computed the TH 
distributions in the steady-states of the BRMPU as well as in 
the models Y and Z through computer simulations. Comparison of 
these distributions are shown in section III. Such comparisons
elucidate the connection between the BRMPU 
and the NaSch model. Then, approximating the BRMPU as a generalization 
of the NaSch model with a {\it time-dependent hopping rate} 
for the buses, we calculate in section IV 
the TH distribution in the BRMPU from 
the corresponding analytical expression in the NaSch model. We 
compare the TH distributions thus derived from analytical 
considerations with the corresponding results of computer 
simulations of the BRMPU. These comparisions do not merely point 
out the regimes of validity of our analytical results but also 
indicate the differences arising from the different natures of 
the low-density steady-states in the BRMPU and the NaSch model. 
Finally, we investigate in section V, 
interesting kinetic phenomena at low 
densities of the BRMPU by computing the appropriate correlation 
functions (to be defined in section V). We extract the 
universal laws governing the growth of the clusters of buses 
in finite samples of BRMPU at low densities where the system 
approaches a Bose-Einstein-like "condensed" state evolving from 
random initial states.

\section{The models and methods}

Let us first summarize  how the totally asymmetric exclusion 
process (TASEP) \cite{sz,gs,spohn}, 
the NaSch model \cite{ns} and the bus route models \cite{evans}
are defined. 

\subsection{TASEP and the NaSch model} 

In the "particle-hopping"  models of traffic the position, speed, 
acceleration as well as time are treated as {\it discrete} variables. 
In this approach, a lane is represented by a one-dimensional lattice. 
Each lattice site represents a "cell" which can be either 
empty or occupied by at most one "vehicle" at a given instant of 
time. At each {\it discrete time} step $t \rightarrow t+1$, the state 
of the system is updated following a well defined prescription. 
In the TASEP a randomly chosen particle can move forward, by one 
lattice spacing, with probability $q$ if the lattice site immediately 
in front of it is empty. In the NaSch model, the speed $v$ of each 
vehicle can take one of the $v_{max}+1$ allowed {\it integer} 
values $v=0,1,...,v_{max}$. If the random-sequential updating scheme 
of the TASEP is replaced by parallel updating then it becomes 
identical to the NaSch model with $v_{max}=1$ and random braking 
probability $p=1-q$. Our interest in the NaSch model is to
unravel its connections to the BRM. For this purpose, we only need
the NaSch model only with $v_{max}=1$. {\it Thus in what follows,
by the NaSch model, we shall mean NaSch model with $v_{max}=1$, 
unless explicitly stated otherwise}. 

\subsection{BRM with parallel and random-sequential updatings}

In the BRM \cite{evans} each of the lattice sites represents a bus 
stop and these stops are labeled by an index $i$ ($i = 1,2,...,L$) 
\cite{evans}. In each step of updating, each bus attempts to hop 
from one stop to the next. Note that in the TASEP and the NaSch model 
one can label the lattice sites by the index $i$ ($i = 1,2,...,L$) 
and describe the state of each of the sites by associating a variable 
$\sigma_i$ with it; $\sigma_i = 1$ if the site $i$ is occupied and 
$\sigma_i = 0$ if the site $i$ is empty. In contrast, in the BRM, 
two binary variables $\sigma_i$ and $\phi_i$ are assigned to each 
site $i$: (i) If the site $i$ is occupied by a bus then $\sigma_i = 1$; 
otherwise $\sigma_i = 0$. (ii) If site $i$ has passengers waiting 
for a bus then $\phi_i = 1$; otherwise $\phi_i = 0$. A site cannot 
have both $\sigma_i = 1$ and $\phi_i = 1$ simultaneously since a 
site cannot have simultaneously a bus and waiting passengers. The 
state of the system is updated according to the following {\it 
random sequential} update rules: a site $i$ is picked up {\it at 
random}. Then, (i) if $\sigma_i = 0$ and $\phi_i = 0$ (i.e, site 
$i$ contains neither a bus nor waiting passengers), then 
$\phi \rightarrow 1$ with probability $\lambda$, where $\lambda$ 
is the probability per unit time of the arrival ($i.e.$ the arrival 
rate) of the first passenger at the empty bus stop. (Arrival of the 
subsequent passengers does not affect the time evolution.)  
(ii) If $\sigma_i = 1$ (i.e., there is a bus at the site $i$) and 
$\sigma_{i+1} = 0$, then the hopping rate $\mu$ of the bus 
from site $i$ to $i+1$ is 
defined as follows: (a) if $\phi_{i+1} = 0$, then $\mu = \alpha$ 
but (b) if $\phi_{i+1} = 1$, then $\mu = \beta$, where $\alpha$ 
is the hopping rate of a bus onto a stop which has no waiting 
passengers and $\beta$ is the hopping rate onto a stop with waiting 
passenger(s). Generally, $\beta < \alpha$, which reflects the fact 
that a bus has to slow down when it has to pick up passengers. In 
the BRMRSU one can set $\alpha = 1$ without loss of generality. 
However, for reasons which will become clear soon, we shall keep 
$\beta < \alpha < 1$. When a bus hops onto a stop $i$ with waiting 
passengers, $\phi_i$ is reset to zero as the bus takes all the 
passengers. Note that the density of buses $c = N/L$ is a {\it 
conserved} quantity whereas that of the passengers is not.

In the BRMPU the random sequential update rules of the BRMRSU are 
replaced by {\it parallel updating} but all the other aspects of 
the updating remain unchanged. BRMPU is related to the NaSch model
in two extreme limits of $\lambda$: In the unphysical limit of 
$\lambda = 0$ (which means the passengers never arrive at a busstop), 
the BRMPU reduces to the NaSch model with $v_{max}=1$ and 
$q = 1-p = \alpha$. In the opposite limit of maximum value of $\lambda$
in BRMPU, $\lambda = \infty$ (very fast rate of passenger arrival at a busstop), 
the BRMPU is equivalent to the NaSch model with $v_{max}=1$ and 
$q = 1-p = \beta$. Note that since the time between two
updating steps is the unit of time, all values of $\lambda \ge 1$
are synonymous with $\lambda = \infty$. This is because any value of
$\lambda \ge 1$ will bring at least one passenger to an empty bus stop
between two updating times. Interesting results in the model occur only for
values of $\lambda \ll 1$.

Note that if we take 
$\alpha = 1$, then the limit $\lambda = 0$ would correspond to 
the limit $q = 1$ (i.e., $p = 0$) of the NaSch model which is a 
deterministic CA and does not exhibit jammed states \cite{nh}. 
Since we are interested in exploring the connection between the 
BRMPU and the NaSch model with arbitrary $q$ throughout this paper 
we consider $\alpha < 1$.

\subsection{Extended NaSch models} 

It has been realized over the last few years that
different modifications of the braking rule in the
NaSch model can lead to different types of phenomena
which are interesting from the perspective of statistical physics.
For example, such modifications can lead to self-organized criticality 
\cite{paczus} as well as metastability and phase segregation
\cite{tt,bjh}. Klauck and Schadschneider \cite{klauck}
considered a model where the particle is allowed to
hop forward by one site or by two sites with two
different hopping rates. It has also been
established that assigning quenched random hopping
rates can lead to the formation of clusters
of vehicles \cite{evansbe,krug,ktitarev}.

In a similar vein, 
we now extend the NaSch model by replacing its constant (time-independent) 
hopping rate $q$ by two other alternatives which are 
intended to mimic the situations in the BRMPU. 

In one of these two alternatives 
(from now onwards referred to as Model Y) the hopping rate 
of a vehicle at a given site $x$ is given by
\begin{equation}
q_x = \beta + (\alpha-\beta)e^{-\Lambda T_{x+1}} 
\label{eq-appqt}
\end{equation}
where $\Lambda > 0$ is a constant and $T_{x+1}$ 
is the time interval that has elapsed since 
the leading vehicle ({\bf LV}) left the site $x+1$. 
$T_{x+1}$ is therefore the time interval between the departure of LV
and the arrival of the following vehicle ({\bf FV}) at the site $x+1$.

In the other extended NaSch model 
(from now onwards referred to as Model Z) the hopping rate 
of the $n$-th vehicle depends on its instantaneous 
DH $\Delta x_n$ :
\begin{equation}
q_n = \beta + (\alpha-\beta)e^{-\Lambda~\Delta x_n/\beta} 
\label{eq-appqx}
\end{equation}
where $\beta$ ($<1$) is a constant. At first sight it may seem 
more appropriate to have $v$, rather than $\beta$, in the 
exponential in equation (\ref{eq-appqx}). However, 
in section III and fig.3(b), we show that the extended NaSch model with the 
hopping rates of the form (\ref{eq-appqx}) is a good 
approximation of the BRMPU in a wide range of circumstances. 

Both the models
Y and Z reduce to the NaSch model with constant
hopping rates $\alpha$ in the limit $\Lambda = 0$,
and reduce to the NaSch model with
a constant hopping rate $\beta$ in the
limit $\Lambda = \infty$. 

Models Y and Z are devised to capture the essential features
of a bus-route model where the time- /space-
dependent hopping rates of the vehicles depend
on the presence or absence of waiting passengers.

\subsection{Methods of Simulation}

For the numerical calculations of the various quantities through 
computer simulations, we let the system evolve from a random initial 
state following the appropriate updating rules mentioned above. 
We compute the quantities relevant for the investigation of the 
kinetics of the system during the time-evolution of the system 
towards its steady-state. After the system reaches steady-state, 
we compute its steady-state properties, e.g, the TH distribution, 
by letting it evolve for the next $5 \times 10^4$ time steps to 
obtain the required data. We then repeat the calculation with 
a different random initial state and, finally, average the data 
over $100$ different random initial states of the system.

The largest systems we have simulated have a total length $L = 10^5$; 
each sample of these was allowed to evolve upto a maximum of 
$10^6$ time steps which is not long enough to reach the corresponding 
steady-state but were used for the study of the kinetics. For the 
computation of the average steady-state properties we have used 
smaller systems (typically $L = 10^4$) which require shorter 
time to reach steady-state.  In all our simulations we have used
a periodic boundary condition.

\section{Results of the extended NaSch models Y and Z} 

\begin{figure}[hbt]
\epsfxsize=3in
\epsfysize=5in
\epsfbox{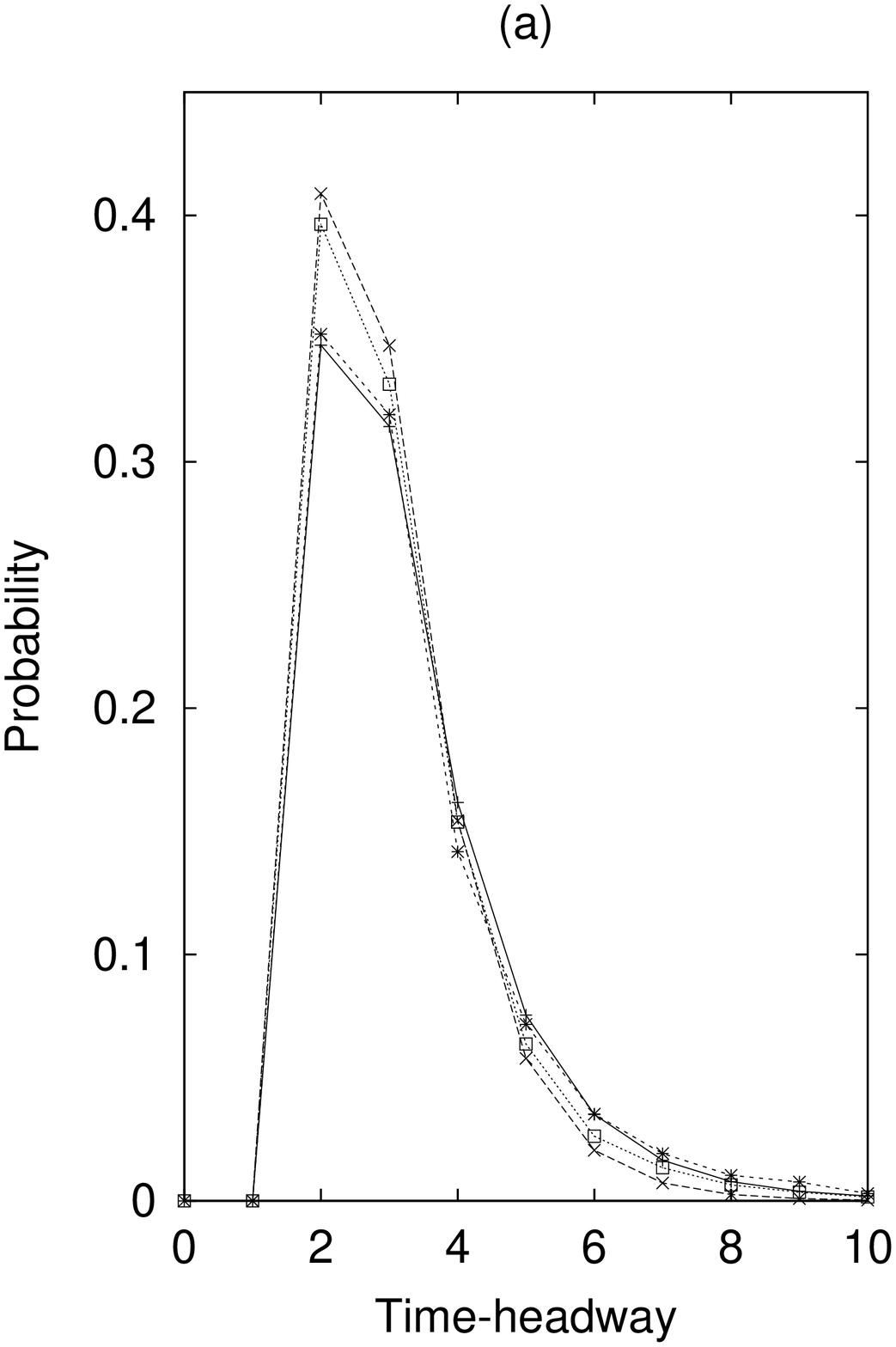}
\vspace*{-5.0in}
\hspace*{3.2in}
\epsfxsize=3in
\epsfysize=5in
\epsfbox{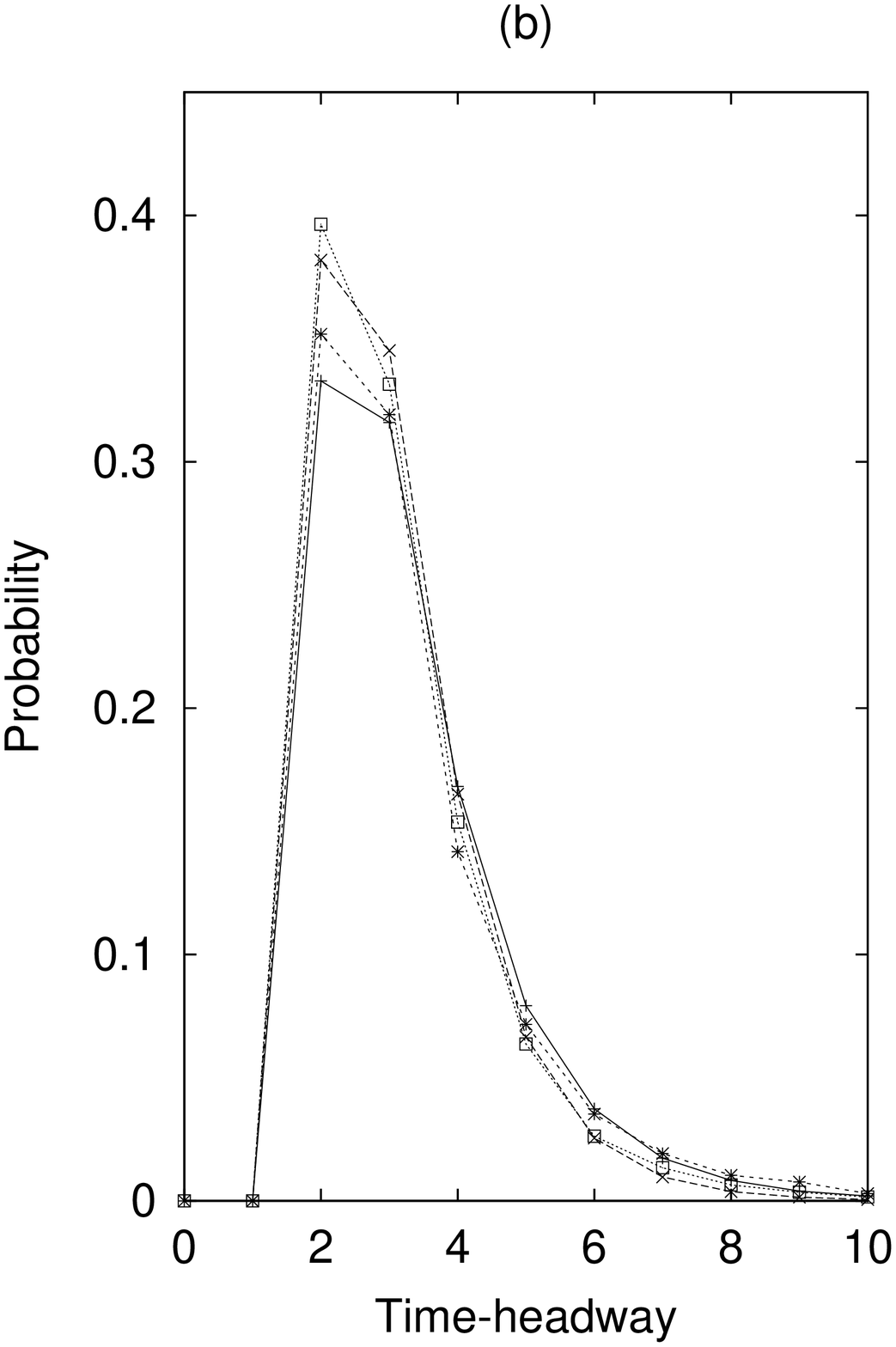}
\vspace{1cm}
\caption{The TH distributions in the (a) model Y, and (b) model Z for 
densities $c = 0.1 (+)$ and $c = 0.5 (\times)$ are compared  against the 
corresponding distributions in the BRMPU for $c = 0.1 (\ast)$ and 
$c = 0.5 (\Box)$. The common parameters are $\Lambda = \lambda = 0.01$, $\alpha = 0.9$
and $\beta = 0.5$. The discrete symbols denote the numerical data obtained 
through computer simulations while the continuous lines joining
these points serve merely as guides to the eye.}
\label{thbrmpu}
\end{figure}

In fig.\ref{thbrmpu} we plot the TH distributions in the models 
Y and Z for $\Lambda = 0.01$, $\alpha = 0.9$, $\beta = 0.5$ at two 
different densities, namely, $c = 0.1$ and $c = 0.5$. Note that
in the special case $\Lambda = 0$, both the models Y and Z as 
well as the BRMPU reduce to the NaSch model with $q = \alpha$. 
The data in fig.\ref{thbrmpu} establish that, when $\Lambda$ is 
sufficiently small (e.g., $\Lambda = 0.01$), the results of the 
models Y and Z agree well with those of the BRMPU at all densities 
for identical values of the set of parameters. 

\begin{figure}[hbt]
\epsfxsize=6in
\epsfysize=5in
\epsfbox{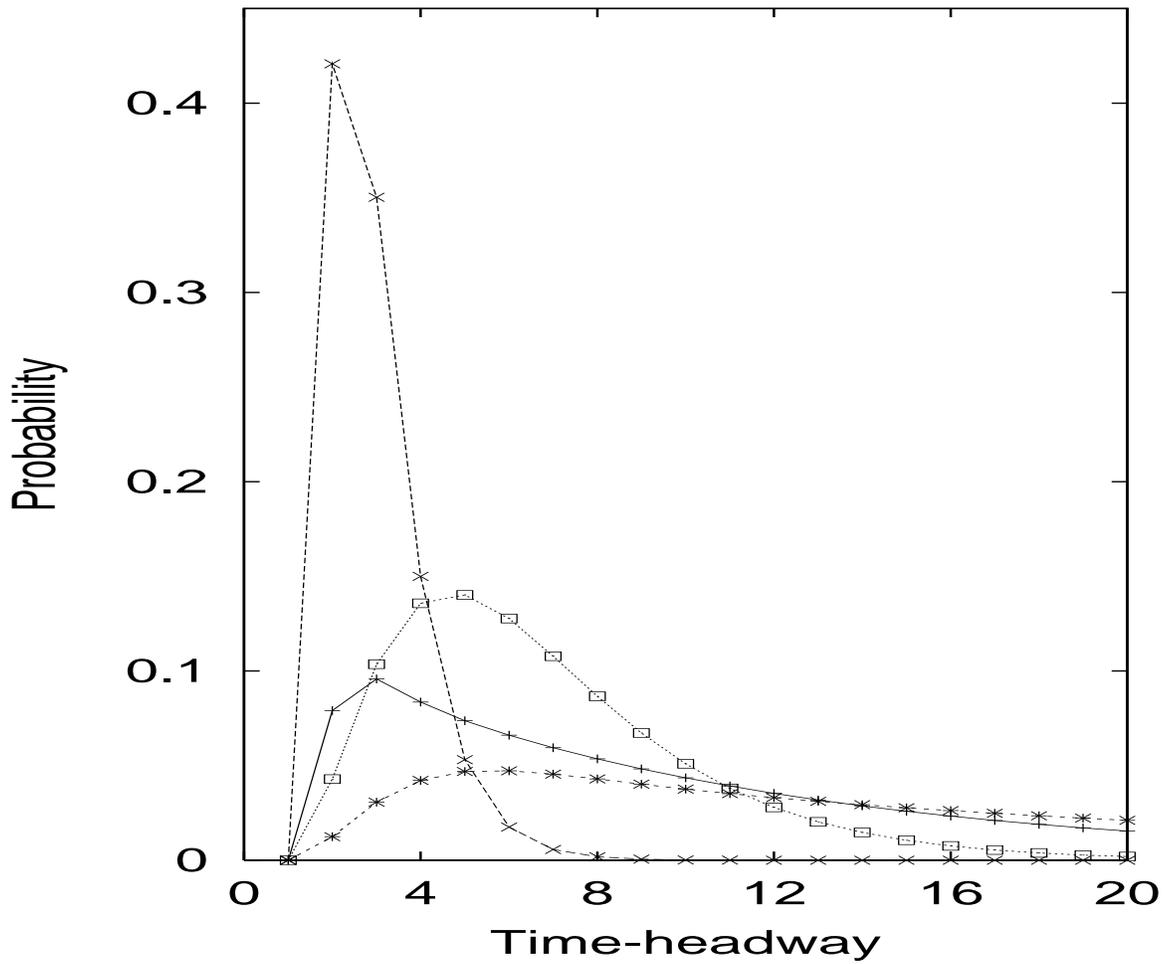}
\vspace{1cm}
\caption{The TH distributions in the NaSch model for 
$q = 0.9, c = 0.5 (\times)$; $q = 0.9, c = 0.1 (+)$;
$q = 0.5, c = 0.5 (\Box)$ and 
$q = 0.5, c = 0.1 (\ast)$.
The discrete symbols denote the exact results obtained
analytically while the continuous lines joining these points
serve merely as guides to the eye.}
\label{thns}
\end{figure}

In order to emphasize the effects of time-/space-dependence of 
the hopping rates on the TH distribution we plot in fig.
\ref{thns} the exact TH distributions in the NaSch model with 
$v_{max} = 1$ for $q = 0.9$ and $q = 0.5$ at the same two 
densities as used in fig.\ref{thbrmpu}. The observation that the 
TH distribution in the NaSch model for $q = 0.9, c = 0.5$ is narrow 
can be explained by fact that small noise ($p = 0.1$) gives rise 
to only a small width of the $\delta$-function-like TH distribution, 
centered at $2$, that one would observe at $c = 0.5$ in the 
deterministic limit $q = 1.0$ of the NaSch model. Comparing fig.\ref{thns} 
with the fig.\ref{thbrmpu} we find that, except for $q = 0.9, c = 0.1$, 
the TH distributions in the NaSch model have much longer tail than 
those in the BRMPU as well as in the model Y and model Z for the 
parameters $\alpha = 0.9, \beta = 0.5$. 

\begin{figure}[hbt]
\epsfxsize=3in
\epsfysize=5in
\epsfbox{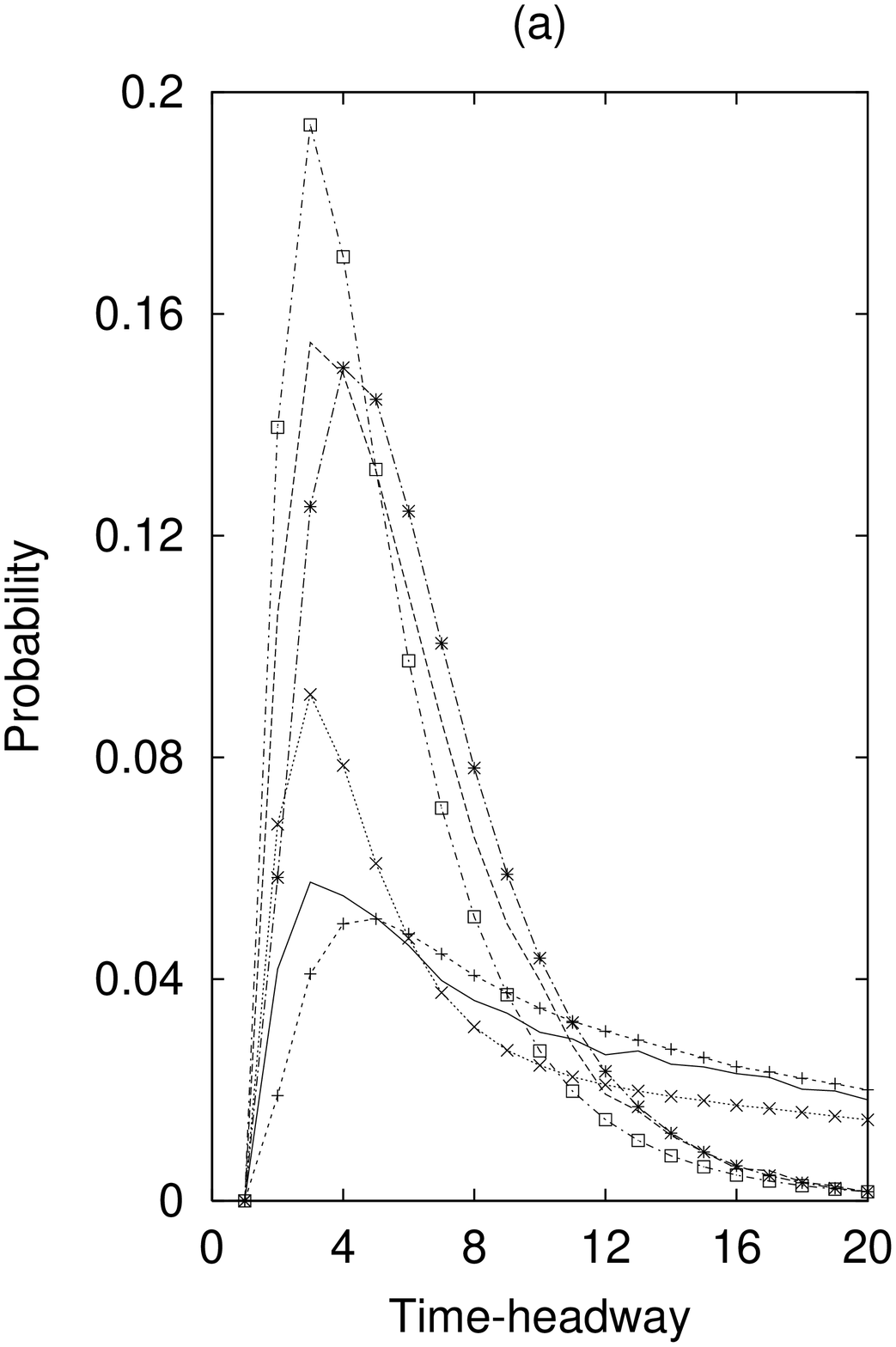}
\vspace*{-5.0in}
\hspace*{3.2in}
\epsfxsize=3in
\epsfysize=5in
\epsfbox{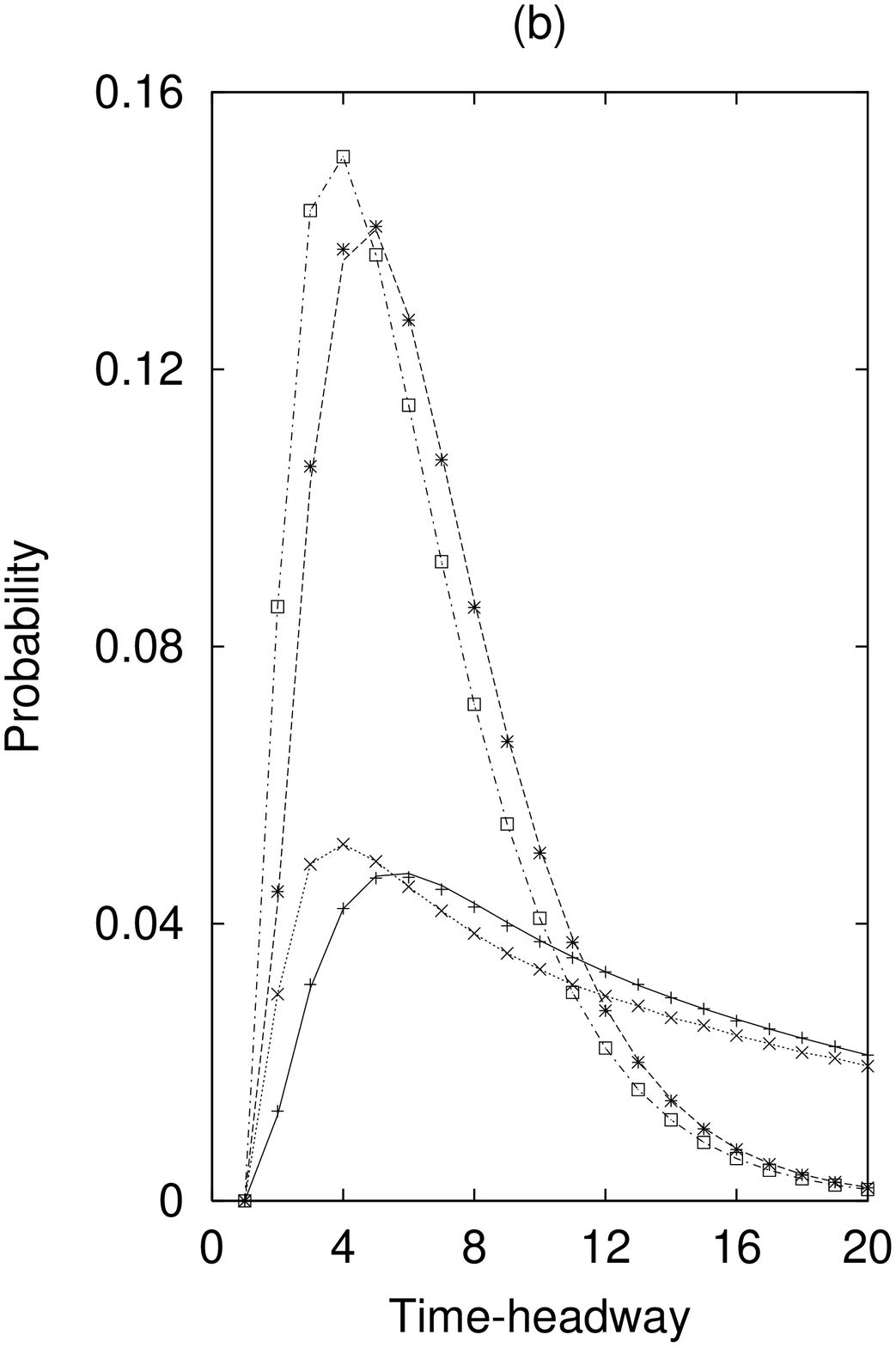}
\vspace{1cm}
\caption{The TH distributions in BRMPU, model Y and model Z 
for (a) $\Lambda = \lambda = 0.5$ and (b) $\Lambda = \lambda = 1.0$;
the common parameters being $\alpha = 0.9, \beta = 0.5$. The
continuous line and the dotted line correspond to $c = 0.1$ and
$c = 0.5$, respectively, in BRMPU, while the discrete data points
denoted by the symbols $+$ and $\ast$ correspond to $c = 0.1$
and $c = 0.5$, respectively, in the model Z. The symbols
$\times$ (connected by dashed line) and $\Box$ (connected by
dashed-dotted line) correspond to $c = 0.1$ and $c = 0.5$,
respectively, in the model Y.}
\label{thyzvsns}
\end{figure}

Thus, the BRMPU is well approximated by both the models Y and Z 
at $\Lambda$ as small as $0.01$. However, we find a larger 
difference between the TH distributions in the models Y and Z 
at larger values of $\Lambda$ (see fig.(\ref{thyzvsns}a)). 
For $\lambda \ge 1$ the BRMPU reduces to the NaSch model with 
$q = \beta$ and the corresponding TH distribution is in excellent 
agreement with that in the model Z but differs significantly 
from that in the model Y (see fig.(\ref{thyzvsns}b)). 

The results in figures 1,2 and 3 show that in a certain density
regime, the BRMPU is well approximated by the
NaSch model with time-/space-dependent hopping rates. 
We expect the passenger arrival rate $\lambda$ of the BRMPU
and the hopping rate $\Lambda$ in models Y and Z to be related.
We assume that the two are equal, even though we continue to use
the two different symbols in order to allow the possibility of
a difference between them in future simulations.

\section{Analytical results of the BRMPU} 

For analytical calculation of the TH distributions 
we label the site (i.e., the bus stop) where the detector is 
located by $j=0$, the stop immediately in front of it by $j=1$, 
and so on. The detector clock resets to $\tau=0$ everytime a bus 
leaves the detector site. We begin our analytical calculations 
by writing ${\cal P}_{th}(\tau)$, the probability of a TH $\tau$ 
between the LV and the FV of a pair, as 
\begin{equation}
{\cal P}_{th}(\tau) = \sum_{t_1=1}^{\tau-1} P(t_1)Q(\tau-t_1|t_1) 
\label{eq-pth}
\end{equation}
where $P(t_1)$ is the probability that there is a time interval $t_1$ 
between the departure of the LV and the arrival of the FV at the 
detector site and $Q(\tau-t_1|t_1)$ is the conditional probability 
given that the FV arrives at the detector site $t_1$ time steps 
after the departure of the LV, it halts for $\tau - t_1$ time steps 
at that site.

Encouraged by the success of the models
Y and Z in capturing the TH distributions over moderate and high
density regimes, we now approximate the BRMPU as an extended NaSch
model with a time-dependent hopping rate which is closely related to 
(but slightly different from) those in the models Y and Z.
Thus for our analytical calculation of ${\cal P}_{th}(\tau)$ in the BRMPU, we 
approximately treat it as an extended NaSch model (with $v_{max} = 1$) where
the hopping rate $q$, instead of being a constant, is a 
time-dependent quantity given by the expression 
\begin{equation}
q = \beta + (\alpha-\beta)e^{-\Lambda t_1}.
\label{eq-appqt1}
\end{equation}
This form can be compared to those given in equations (\ref{eq-appqt}) and
(\ref{eq-appqx}).

The exact analytical expression for $P(t_1)$ in the NaSch model (with 
$v_{max} = 1$) has been derived earlier \cite{ghosh,chowepjb} using 
a 2-cluster approximation \cite{ssni} which goes beyond the simple
mean field approximation. Following the same arguments we now get 
\begin{equation}
P_{cl}(t_1)  =  {\cal C}(1|\underline{0})q
        \left[{\cal C}(0|\underline{0})q + p \right]^{t_1-1}
\label{eq-clpt1}
\end{equation}
where $q$ is given by (\ref{eq-appqt1}) and ${\cal C}$ gives the 
2-cluster steady-state configurational probability for the argument 
configuration; the underline under an argument of ${\cal C}$ implies 
the associated condition. The expressions for the various ${\cal C}$s
are given by \cite{ssni,ghosh,chowepjb}
\begin{equation}
{\cal C}(\underline{1}|0) \quad =\quad {\cal C}(0|\underline{1}) = \frac{y}{c}
\label{eq-c1}
\end{equation}
\begin{equation}
{\cal C}(\underline{0}|1) \quad =\quad {\cal C}(1|\underline{0}) =  \frac{y}{d}
\label{eq-c2}
\end{equation}
\begin{equation}
{\cal C}(\underline{1}|1) \quad =\quad {\cal C}(1|\underline{1}) = 1 - \frac{y}{
c}
\label{eq-c3}
\end{equation}
\begin{equation}
{\cal C}(\underline{0}|0) \quad =\quad {\cal C}(0|\underline{0}) = 1 - \frac{y}{ d}
\label{eq-c4}
\end{equation}
where
\begin{equation}
y = \frac{1}{2q}\left( 1 - \sqrt{1 - 4 q c d}\right),
\label{eq-y}
\end{equation}
$q = 1 - p$ and $d = 1 - c$.

On the other hand, in the simple mean field approximation, the 
2-cluster probabilities reduce to 
${\cal C}(1|\underline{0}) \rightarrow c$ and 
${\cal C}(0|\underline{0}) \rightarrow 1-c$ and, hence,
\begin{equation}
P_{mf}(t_1)  =  c q [1 - c q]^{t_1-1} 
\label{eq-mfpt1}
\end{equation}
We shall calculate the TH distribution, ${\cal P}_{th}(\tau)$, 
given in (\ref{eq-pth}), 
using the expression (\ref{eq-clpt1}) 
(together with (\ref{eq-c2}), (\ref{eq-c4}) and (\ref{eq-y})) and then 
compare with the corresponding TH distribution obtained by using 
(\ref{eq-mfpt1}), instead of (\ref{eq-clpt1}), to emphasize the 
importance of correlations. 

In order to obtain ${\cal P}_{th}(\tau)$, let us next calculate $Q(\tau-t_1\mid t_1)$. 
Again, following the 
arguments used earlier \cite{chowepjb} in the calculation of the TH 
distribution in the NaSch model, we get 
\begin{eqnarray}
Q(\tau-t_1\mid t_1) &=& (1 - \bar{g}^{t_1})p^{\tau-t_1-1}q \nonumber \\ 
& & + \bar{g}^{t_1}gq \frac{[(\bar{g})^{\tau-t_1-1} - (p)^{\tau-t_1-1}]}{\bar{g}-p} 
\label{eq-q}
\end{eqnarray}
where $g$ is the probability that a vehicle moves in the next time 
step (i.e., in the $(t+1)^{th}$ time step) and $\bar{g} = 1-g$. In 
the 2-cluster approximation 
\begin{equation}
g_{cl} = q  {\cal C}(\underline{1}|0)
\label{eq-clg}
\end{equation}
which, in the simple mean field approximation reduces to
\begin{equation}
g_{mf} = q (1-c) 
\label{eq-mfg}
\end{equation}

Substituting (\ref{eq-clpt1}) and (\ref{eq-q}) into (\ref{eq-pth}) 
and using (\ref{eq-clg}) for $g$ and (\ref{eq-appqt1}) for $q$  
(together with (\ref{eq-c1})-(\ref{eq-c4}) for the configurational 
probabilities and (\ref{eq-y})), we get ${\cal P}_{th}(\tau)$ in the 2-cluster 
approximation by carrying out the summation over $t_1$ in 
(\ref{eq-pth}) numerically. We shall refer to this result as the 
2-cluster estimate of the TH distribution. Similarly, substituting 
(\ref{eq-mfpt1}) and (\ref{eq-q}) into (\ref{eq-pth}) and using 
(\ref{eq-mfg}) for $g$ and (\ref{eq-appqt1}) for $q$ (together with 
(\ref{eq-c1})-(\ref{eq-c4}) and (\ref{eq-y})) we get the simple
mean field estimate of ${\cal P}_{th}(\tau)$ by again summing over 
$t_1$ numerically. Note that in both cases, 
${\cal P}_{th}(\tau)$, in addition to
its $\tau$ dependence, depends on parameters $c, \alpha, \beta, 
\Lambda.$

\begin{figure}[hbt]
\epsfxsize=3in
\epsfysize=5in
\epsfbox{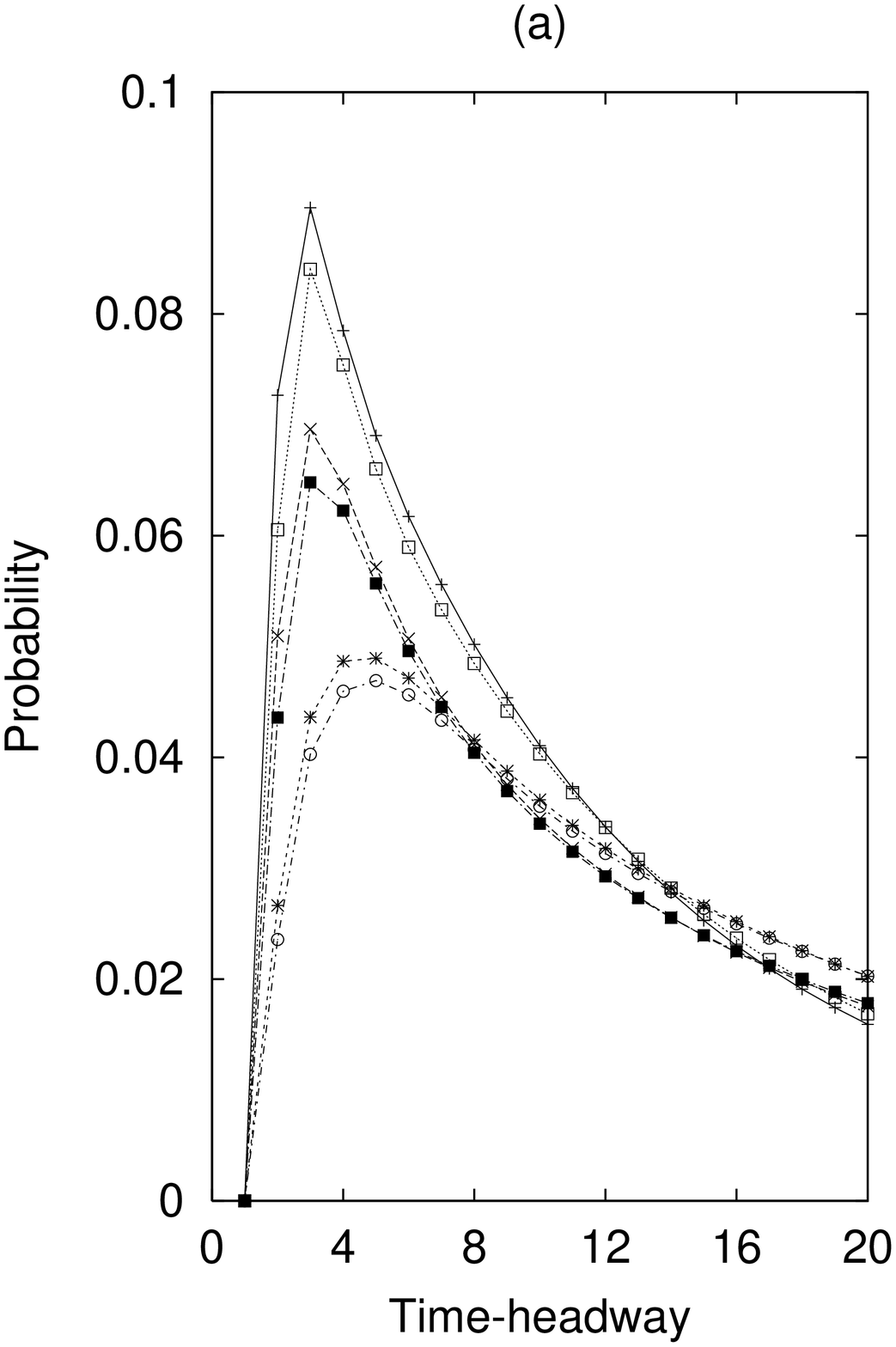}
\vspace*{-5.0in}
\hspace*{3.2in}
\epsfxsize=3in
\epsfysize=5in
\epsfbox{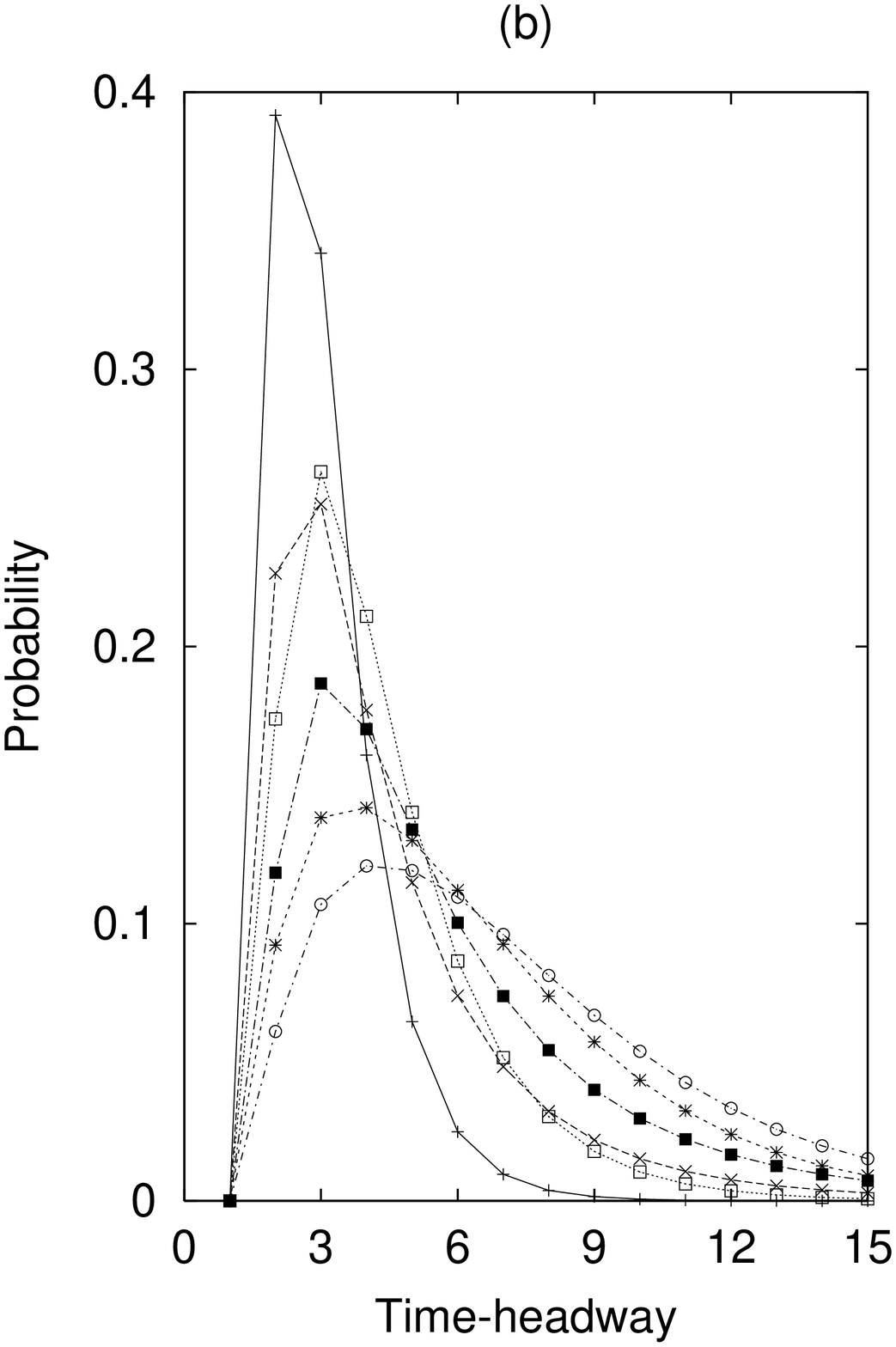}
\vspace{1cm}
\caption{The TH distributions in the BRMPU calculated analytically
at densities (a) $c = 0.1$ (b) $c = 0.5$.
The symbols $+ , \times, (\ast)$ correspond to $\Lambda = 0.01, 0.10, 0.50$,
respectively, in the 2-cluster approximation, while the symbols
$\Box, \protect\rule{2mm}{2mm}\protect, \circ$ correspond to 
$\Lambda = 0.01, 0.10, 0.50$,
respectively, in the simple mean field approximation.
The common parameters are $\alpha = 0.9$ and $\beta = 0.5$. The discrete
symbols denote the results obtained through analytic eq.(\ref{eq-pth}), 
while the continuous lines joining these points
serve merely as guides to the eye. Note also the different scales along
both the axes in (a) and (b).}
\label{cvecv}
\end{figure}

In fig.\ref{cvecv} we show these analytic results for three different
values of $\Lambda$ and two different values of densities $c$.
At sufficiently low density of buses, there is hardly any difference 
between the 2-cluster estimate and the simple mean field estimate 
of the TH distribution (see fig.\ref{cvecv}(a)). However, with the 
increase of the density of the buses, the difference between these 
two estimates increases (see fig.\ref{cvecv}(b)). 

As noted earlier, the TH distribution in the 
BRMPU changes continuously with the variation of $\lambda$; the 
results for $\lambda = 0$ and $\lambda = 1$ are identical to those 
in the NaSch model with $q = \alpha$ and $q = \beta$, 
respectively. 
 
\begin{figure}[hbt]
\epsfxsize=3in
\epsfysize=5in
\epsfbox{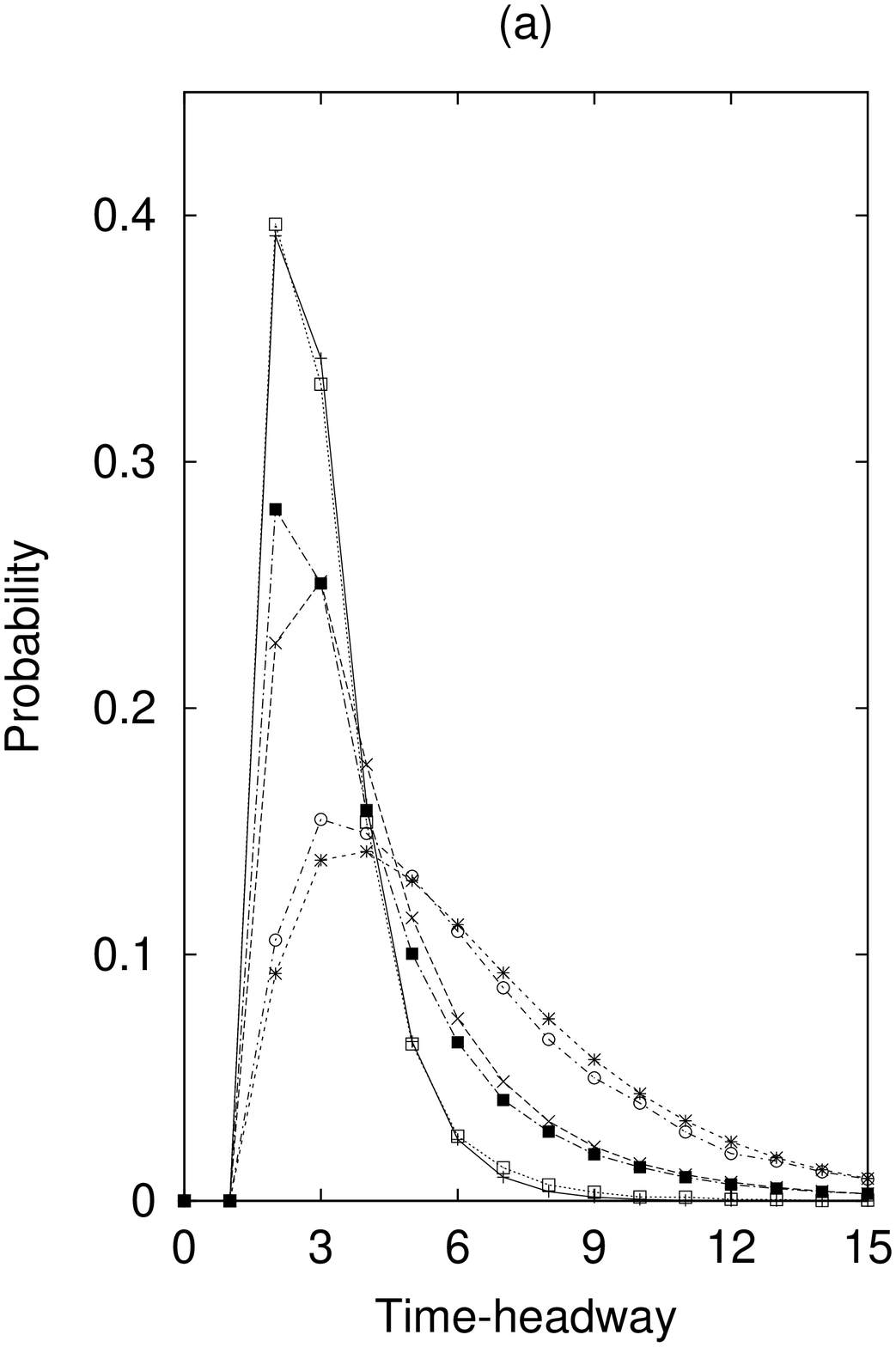}
\vspace*{-5.0in}
\hspace*{3.2in}
\epsfxsize=3in
\epsfysize=5in
\epsfbox{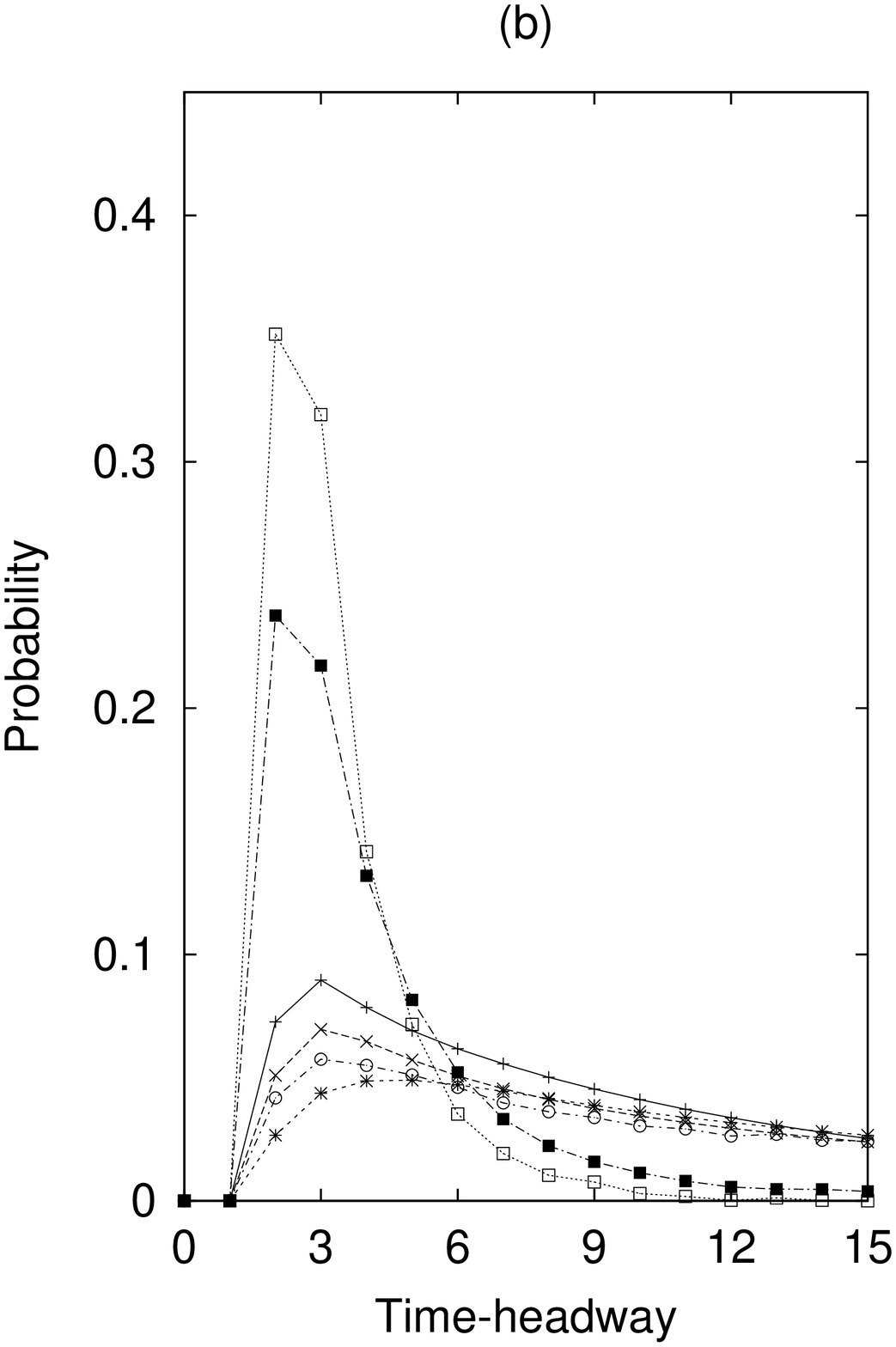}
\vspace{1cm}
\caption{The TH distributions in the BRMPU calculated analytically
at densities (a) $c = 0.5$ (b) $c = 0.1$.
The symbols $+ , \times, (\ast)$ correspond to the TH distributions
in the BRMPU, obtained analytically, for $\Lambda = 0.01, 0.10, 0.50$,
respectively, in the 2-cluster approximation while the symbols
$\Box, \protect\rule{2mm}{2mm}\protect, \circ$ correspond to the 
TH distributions
in the BRMPU, obtained through computer simulations, for
$\lambda = 0.01, 0.10, 0.50$, respectively.
The common parameters are $\alpha = 0.9$ and $\beta = 0.5$.
The continuous lines joining these points
serve merely as guides to the eye. Note also the different scales along
both the axes in (a) and (b).}
\label{ana}
\end{figure}

In fig. \ref{ana} we compare the 2-cluster analytic estimate 
of the TH distribution in the BRMPU (approximated as the extended
NaSch model) for three different  
values of $\Lambda$, namely, $\Lambda = 0.01, 0.10, 0.50$, with the 
corresponding numerical data we have obtained from direct computer 
simulations of the BRMPU model. 
Figure  \ref{ana} (a) shows the comparison for the higher density value $c=0.5$.
We note very good agreement between the 2-cluster 
estimates and the computer simulation data.
Similar comparison at a lower density $c=0.1$ is shown in fig. \ref{ana} (b).
The poor agreement between the 2-cluster estimate and simulation 
data for the TH distribution in the BRMPU at low densities is a 
consequence of the fact that at low densities, the vehicles 
in a finite system form a cluster in the steady-state where there 
is at most one or two empty sites in between each pair of vehicles. 
It is well known that the 2-cluster approximation scheme is not 
good enough for such states where correlation extends over  
distances which are much longer than what can be captured by a 
2-cluster approximation. On the other hand, at higher densities 
there is no clustering of the buses in the steady-state of the BRMPU 
and the physics of the system is very similar to that in the 
steady-states of the NaSch model. This is because most of the buses
stop on account of another bus at the next site, rather than due to
waiting passengers.

Nevertheless, since the states with clusters of buses are metastable 
in infinitely long samples of BRMPU, the 2-cluster approximation is 
expected to yield good estimates for the stable steady-states of 
the BRMPU even at low densities. However, since it is extremely 
difficult (requires very long simulation time)
to achieve these stable steady-states in any computer simulation at 
small $\lambda$ we have not been able to demonstrate this explicitly.

\section{Kinetics in the BRMPU} 

In traffic models like BRMPU, the kinetics are governed by two coupled
fields, local passenger density $\phi_i (t)$ and local bus density 
$\sigma_i (t)$. In this paper a binary approximation (zero or nonzero)
to the $\phi_i (t)$ field is made. $\sigma_i (t)$ is globally 
conserved and $\phi_i (t)$ is a nonconserved field. It is not clear
whether the kinetics seen in our simulations of BRMPU and models Y and Z 
are derivable from a free energy functional. We are currently 
inquiring into such a possibility.
If this turns out to be the case, then the model 
of kinetics appropriate to our simulations is model C \cite{modelc} in the
Halperin- Hohenberg classification scheme of critical dynamics.

Let us define the correlation function 
\begin{equation}
{\cal C}(r,t) = \biggl[\frac{1}{L} \sum_{i=1}^L \sigma_i(t) \sigma_{i+r}(t) - c^2 \biggr] 
\label{eq-cf}
\end{equation}
where $t = 0$ corresponds to the initial state. The symbol 
$\biggl[.\biggr]$ indicates average over random initial 
conditions. By definition, ${\cal C}(r,t)$ vanishes in the absence 
of any correlation in the occupation of the sites by the buses. 
Also at any time $t$, ${\cal C}(r=0,t) = c(1-c).$
This correlation function has been calculated earlier for the NaSch 
model analytically for $v_{max} = 1$ \cite{as99} and numerically 
for higher values of $v_{max}$. However, the nature of this 
correlation function in the BRMPU is expected to differ 
qualitatively, particularly at low densities, from that in the 
NaSch model because of the formation of clusters in the steady-state 
of the BRMPU.

\begin{figure}[hbt]
\epsfxsize=6in
\epsfysize=5in
\epsfbox{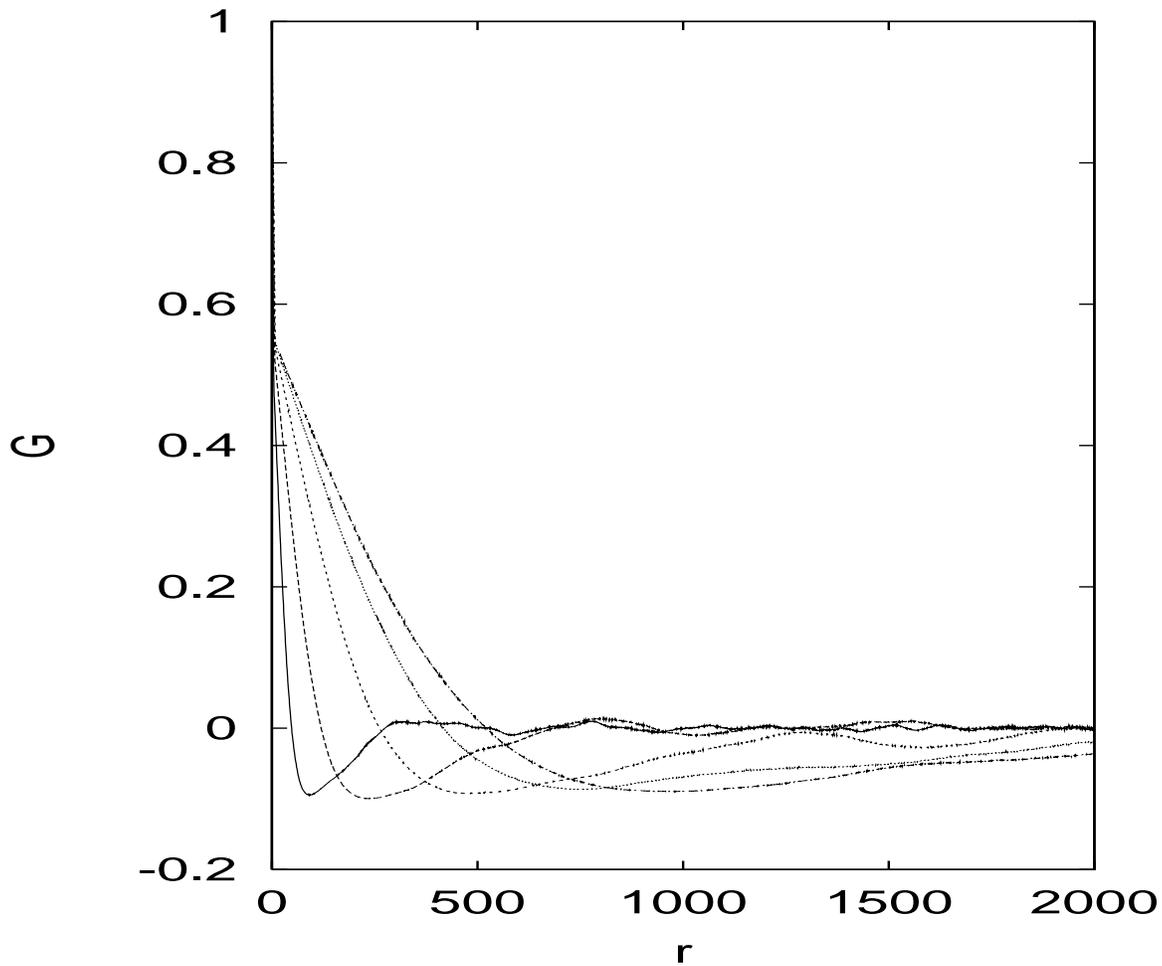}
\vspace{1cm}
\caption{The normalized correlation function $G(r,t)$ in the BRMPU plotted
against $r$ for (from left to right) $t = 5 \times 10^3$ (solid line), 
$t = 10^5$ (long dashes),
$t = 10^6$ (short dashes), $t = 3 \times 10^6$ (dots), $t = 5 \times 10^6$
(dashed-dotted line) at a density $c = 0.1$. The values of the other
parameters are  $\lambda = 0.01, \alpha = 0.9, \beta = 0.5$. }
\label{cffig}
\end{figure}

We compute ${\cal C}(r,t)$ during the time-evolution of the system 
from random initial states. Our simulations for the time evolution 
are done only for one set of parameters: $c = 0.1, \lambda = 0.01, 
\alpha = 0.9$ and $\beta = 0.5.$ In fig.\ref{cffig} we plot the 
{\it normalized} correlation function 
\begin{equation}
G(r,t) = {\cal C}(r,t)/{\cal C}(r=0, t) = {\cal C}(r,t)/(c(1-c)) 
\end{equation}
as a function of $r$ for values of $t$ upto $5 \times 10^6$. 

\begin{figure}[hbt]
\epsfxsize=6in
\epsfysize=5in
\epsfbox{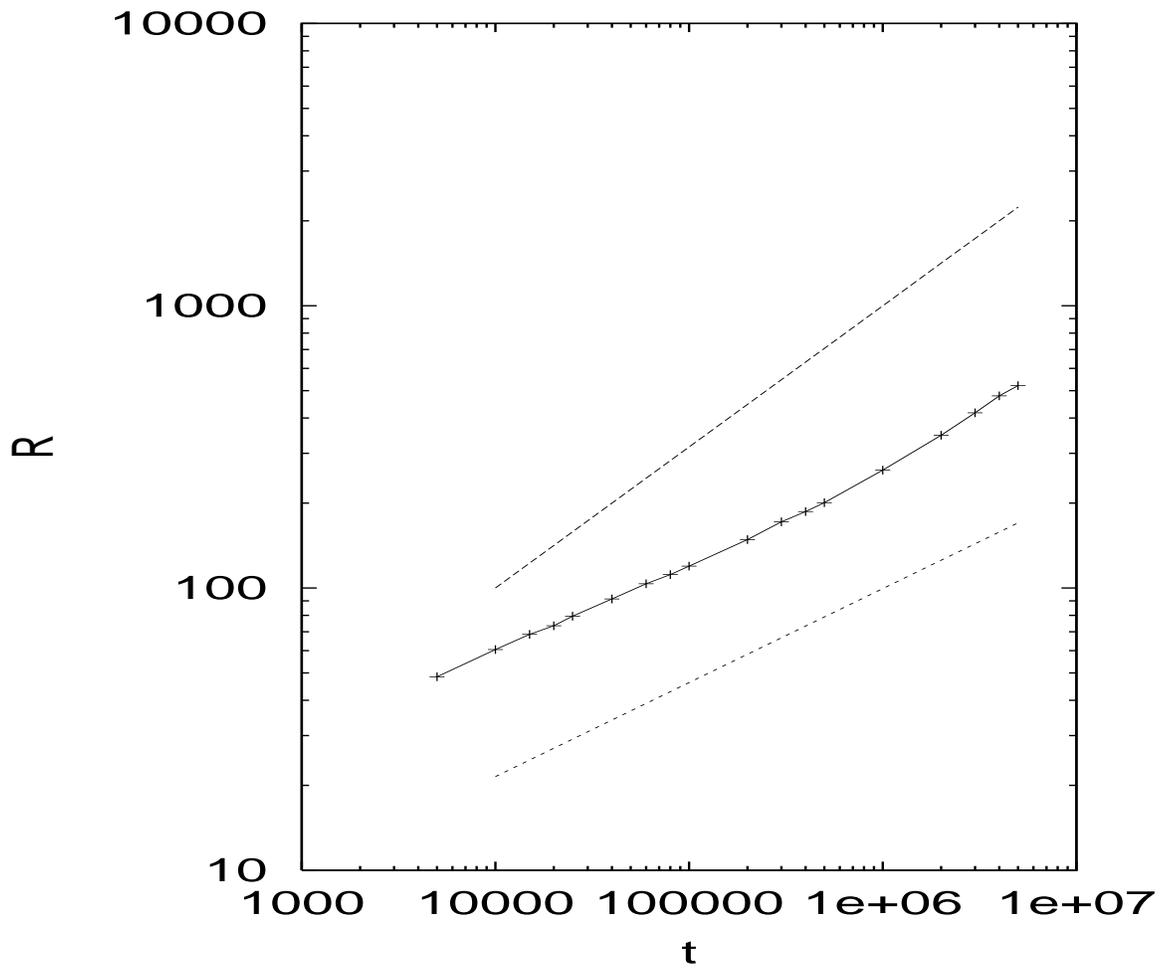}
\vspace{1cm}
\caption{The log-log plot of $R(t)$ in the BRMPU at $c = 0.1$.
The values of the other parameters are
$\lambda = 0.01, \alpha = 0.9, \beta = 0.5$. The dashed (top) and dotted 
(bottom) lines have slopes of $1/2$ and $1/3$ respectively.}
\label{loglogr}
\end{figure}

\begin{figure}[hbt]
\epsfbox{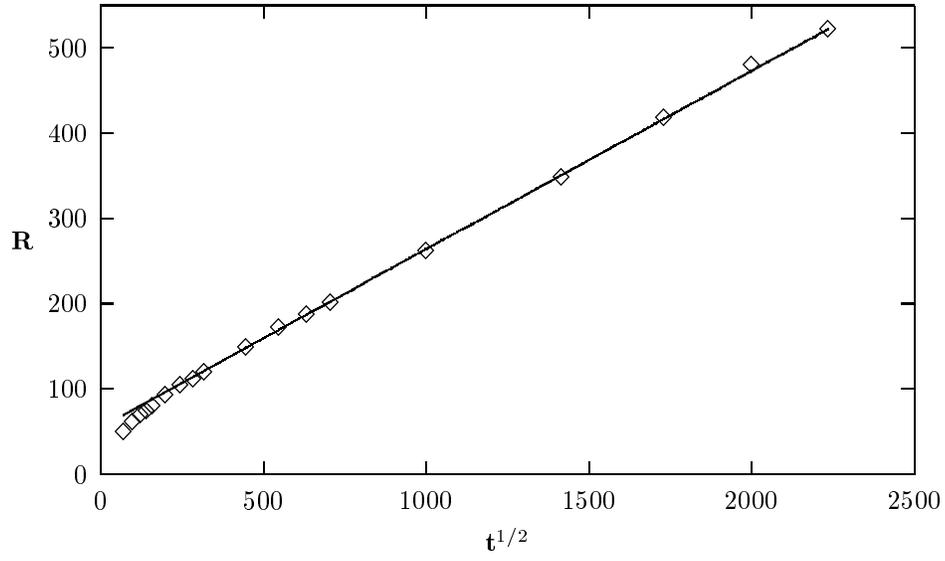}
%\hspace*{-1.5in}
%\epsfxsize=6in
%\epsfysize=5in
%\epsfbox{newfig8.ps}
\vspace{-9cm}
\caption{The discrete data points represent $R(t)$ in the BRMPU
at $c = 0.1$ plotted against $t^{1/2}$. The continuous solid line
corresponds to $55 + 0.2 t^{1/2}$. The values of the other parameters
are $\lambda = 0.01, \alpha = 0.9, \beta = 0.5$.}
\label{rvsthalf}
\end{figure}

The value $r = R$ corresponding to the first zero-crossing of 
$G(r,t)$ is taken as a measure of the typical size of the clusters of buses 
at time $t$ \cite{bray}. The fact that $R$ increases with 
$t$ indicates the coarsening of these clusters. It is 
worth mentioning here that in systems with conserved order 
parameters (the so-called model B, of which the binary alloy is 
a physical realization) the coarsening follows the Lifshitz-Slyozov 
law $R(t) \sim t^{1/3}$. In the case of the BRMPU, $R(t)$ may 
appear to follow the same Lifshitz-Slyozov law if the data upto 
$t \simeq 10^6$ are shown on a log-log plot (see fig.\ref{loglogr}). 
However, the upward turn of the data beyond $t \simeq 10^6$ in 
fig.\ref{loglogr} indicates more subtle features of this growth. 
In fact, fitting the raw data $R(t)$ to the curve 
\begin{equation}
R(t) = R_0 + A t^{1/2} 
\label{thalflaw}
\end{equation}
we have estimated the parameters $R_0$ and $A$. We found that 
$R_0 \simeq 55$ and $A \simeq 0.2$. If the growth of $R(t)$, 
indeed, follows the law (\ref{thalflaw}) then the $t^{1/2}$ growth 
law is expected to become clearly visible directly in 
fig.\ref{loglogr} for times long enough to satisfy the condition 
$A t^{1/2} >> R_0$, i.e., for $t >> 10^5$. This argument, together 
with our estimate $R_0 \simeq 55$, explains why the true growth 
law (\ref{thalflaw}) can be anticipated in fig.\ref{loglogr} only 
beyond $t \simeq 10^6$. In fact, plotting $R(t)$ against 
$t^{1/2}$ and comparing with $55 + 0.2 t^{1/2}$ in fig.\ref{rvsthalf} 
we do, indeed, see clear evidence of the $t^{1/2}$ growth for 
$t >> 10^5$.

\begin{figure}[hbt]
\epsfxsize=6in
\epsfysize=5in
\epsfbox{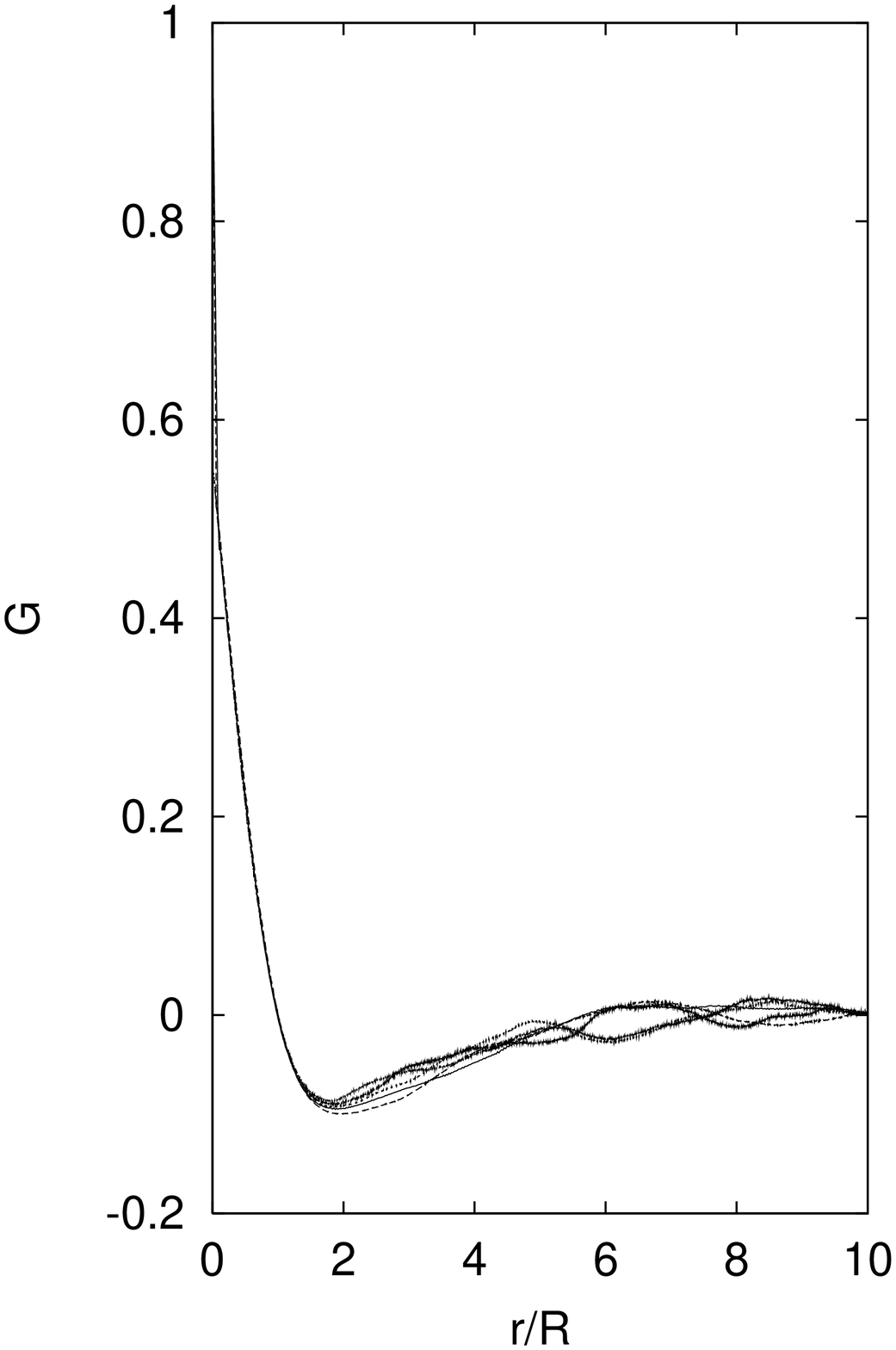}
\vspace{1cm}
\caption{The normalised correlation function $G$ in the BRMPU at 
$c = 0.1$, for the same five values
of $t$ as in the fig.\ref{cffig}, are plotted against $r/R(t)$.
The values of the parameters $\lambda, \alpha$ and $\beta$ are also
identical to those in the fig.\ref{cffig}. }
\label{scaling}
\end{figure}

Finally, in fig.\ref{scaling} we plot the normalized correlation 
function $G$ against the scaled variable $r/R(t)$. Since the 
data for $t$ as widely separated as $t = 5 \times 10^3$ to 
$t = 5 \times 10^6$ superpose, the validity of dynamic 
scaling \cite{bray} is convincing for kinetics of the BRMPU model. 
The $t^{1/2}$ growth in our BRMPU model that is akin to model C 
\cite{modelc} implies that
it is the nonconserved passenger density field $\phi$ that is driving
the kinetics, for the parameter set that we have simulated.

\section{Conclusions}

The models and simulations presented in this paper were inspired 
by the work on BRMRSU in ref. \cite{evans}. The model in 
\cite{evans} uses random sequential updating (RSU), whereas our 
complimentary study is based on parallel updating (PU). As expected,
the properties of the steady states including the dependence of average
velocity and current on the particle density are similar (not 
explicitly displayed in figures). We have in addition computed
and analytically obtained (in the $2$-cluster approximation) the
time headway (TH) distributions for a wide range of parameters.
At moderate and high densities, the simulation and analytic results
agree well; but the comparison fails at low densities. We have also
studied kinetics of the BRMPU which is also complimentary to that done
for the BRMRSU in ref. \cite{evans}. We find the bus clusters to grow
in size as $t^{1/2}$. This shows that the growth exponent is robust
with respect to the updating schemes. We have also computed the equal 
time pair correlation function of the local bus density, and find that
it obeys a dynamical scaling ansatz: $G(r,t) = G(r/R(t))$, where
$R(t)$ is the bus cluster size.

We have also found connection between the models of BRMPU type and the
NaSch model.
The NaSch model is the minimal CA model of vehicular traffic on 
idealized single-lane highways. This model has been extended 
\cite{paczus,tt,bjh,klauck,barlovic,sss2s,wolf}
in various ways to incorporate some aspects of real 
traffic which are not captured by the minimal model. All the bus 
route models which we have considered in this paper, namely, 
the BRMPU, model Y and model Z, may be regarded as extensions 
of the NaSch model with $v_{max} = 1$. In each of these models
the dynamics of the vehicles are coupled to another non-conserved 
field, namely, the passenger density field $\phi$, resulting 
in space- and time-dependent hopping rates of the vehicles. 
However, there is a difference in the length and time scales 
in the bus route models and in the NaSch model. In the NaSch model the 
motion of the vehicles from one cell to the next is described 
on a time scale which is roughly the reaction time of each 
driver. In contrast, the lattice constant in the BRM is of the 
order of the distance between successive stops on the bus route. 
Thus, each time step corresponds to a much longer real
time than that in the NaSch model. In other words, the interactions 
of the buses with the traffic, on its way from one stop to the 
next, are included in the BRM models only through the phenomenological
rate constant $\alpha$. It would be interesting to extend the
bus route models further by including the interaction between a bus and
other vehicles as it moves from one stop to the next.
Inter-vehicle interactions are a natural ingredient of the NaSch model,
and this prescription can be incorporated in the bus route models.

\noindent{\bf Acknowledgements:} We thank A. Schadschneider and D. Stauffer 
for valuable comments and suggestions, and the Department of Physics at 
the University of Toronto for providing us free CPU time on Helios2 
without which this work could not have been completed. This work was 
also supported by NSERC of Canada.

%\newpage 

%\begin{thebibliography}{99}

\newpage

\newpage

\newpage

\newpage

\newpage

\newpage

\newpage

\newpage

\newpage


\begin{references}

\bibitem{sz} B. Schmittmann and R.K.P. Zia, in: {\em Phase Transitions
and Critical Phenomena}, Vol.~17, eds. C. Domb and J.L. Lebowitz
(Academic Press, 1995); Phys. Rep. {\bf 301}, 45 (1998); R.K.P. Zia,
L.B. Shaw, B. Schmittmann and R.J. Astalos, {\tt cond-mat/9906376}.

\bibitem{gs} G. Sch\"utz, in: {\em Phase transitions and critical
phenomena}, eds. C. Domb and J.L. Lebowitz (to appear).

\bibitem{spohn} H. Spohn, {\em Large scale dynamics of interacting
particles} (Springer, 1991).

\bibitem{vp} V. Privman (ed.), {\em Nonequilibrium Statistical
Mechanics in One Dimension} (Cambridge University Press, 1997).

\bibitem{cursci} D. Chowdhury, L. Santen and A. Schadschneider, 
Curr. Sci. {\bf 77}, 411 (1999).

\bibitem{chowpr} D. Chowdhury, L. Santen and A. Schadschneider, preprint
(1999) for Physics Reports.

\bibitem{proc1} D.E. Wolf, M. Schreckenberg and A. Bachem (eds.), {\em
Traffic and Granular Flow} (World Scientific, Singapore, 1996).

\bibitem{proc2} M. Schreckenberg and D.E. Wolf (eds.), {\em Traffic and
Granular Flow '97} (Springer, Singapore, 1998).

\bibitem{wolfram} S. Wolfram, {\em Theory and Applications of Cellular
Automata}, (World Scientific, 1986); {\em Cellular Automata and
Complexity} (Addison-Wesley, 1994).

\bibitem {ns} K. Nagel and M. Schreckenberg, {\it J. Physique I}, {\bf2},
2221 (1992).

\bibitem{ssni} M. Schreckenberg, A. Schadschneider, K. Nagel and N. Ito,
{\it Phys.Rev.E}, {\bf 51}, 2939 (1995).

\bibitem{bml} O. Biham, A.A. Middleton and D. Levine,
Phys. Rev. A {\bf 46}, R6124 (1992).

\bibitem{may}A.D. May, {\sl Traffic Flow Fundamentals} 
(Prentice-Hall, 1990).

\bibitem{leutz} W. Leutzbach, 
{\sl Introduction to the Theory of Traffic Flow}
(Springer, Berlin, 1988).

\bibitem{nagel} K. Nagel, J. Esser and M. Rickert, 
in Annu. Rev. Comp. Phys., volume VIII
ed. D. Stauffer (World Scientific, March 2000).

\bibitem{evans} O.J. O'Loan, M.R. Evans and M.E. Cates, 
Phys. Rev. E {\bf 58}, 1404 (1998).

\bibitem{evansbe} M.R. Evans, Europhys. Lett. Europhys. Lett. 
{\bf 36}, 13 (1996); J. Phys. A {\bf 30}, 5669 (1997).

\bibitem{krug} J. Krug and P. A. Ferrari, J. Phys. A {\bf 29}, 
L465 (1996); see also J. Krug, in: ref.~\cite{proc2}.

\bibitem{ktitarev} D. Ktitarev, D. Chowdhury and D.E. Wolf, J. Phys. A 
{\bf 30}, L221 (1997).

\bibitem{nh} K. Nagel and H.J. Herrmann, Physica A {\bf 199}, 254 (1993).

\bibitem{paczus} K. Nagel and M. Paczuski,
Phys. Rev. E {\bf 51}, 2909 (1995);
see also M. Paczuski and K. Nagel, in: \cite{proc1}.

\bibitem{tt} M. Takayasu and H. Takayasu,
Fractals, {\bf 1}, 860 (1993).

\bibitem{bjh}S.C. Benjamin, N.F. Johnson and P.M. Hui,
J. Phys. A {\bf 29}, 3119 (1996).

\bibitem{klauck} K. Klauck and A. Schadschneider,
Physica A {\bf 271}, 102 (1999)

\bibitem{ghosh}  K. Ghosh, A. Majumdar and D. Chowdhury,
Phys. Rev. E {\bf 58}, 4012 (1998).

\bibitem{chowepjb} D. Chowdhury, A. Pasupathy and S. Sinha,
Eur. Phys. J. B {\bf 5}, 781 (1998).

\bibitem{modelc} C. Sagui, A.M. Somoza and R.C. Desai, 
Phys. Rev. E {\bf 50}, 4865 (1994), and references therein.

\bibitem{as99} A. Schadschneider, Eur. Phys. J. {\bf B10}, 573 (1999).

\bibitem{bray} A. Bray, Adv. in Phys. {\bf 43}, 357 (1994).

\bibitem{barlovic} R. Barlovic, L. Santen, A. Schadschneider
and M. Schreckenberg, Eur. Phys. J. {\bf 5}, 793 (1998).

\bibitem{sss2s} A. Schadschneider and M. Schreckenberg,
Ann. der Phys. {\bf 6}, 541 (1997).

\bibitem{wolf} For generalizations of the NaSch model to multi-lane 
traffic see, for example, K. Nagel, D.E. Wolf, P. Wagner, P. Simon,
Phys. Rev. E {\bf 58}, 1425 (1998) and references therein.
 
%\end{thebibliography}
\end{references}
\end{document}